\documentclass[a4paper,11pt]{article}
\usepackage{jheppub}
\usepackage[T1]{fontenc}
\usepackage{amsfonts}
\usepackage{amsmath}
\usepackage{amssymb,epsf}
\usepackage{latexsym}
\usepackage{graphicx,epsfig}
\usepackage{amssymb}
\usepackage{subcaption}
\usepackage{epstopdf}

\title{Analytical and Numerical Study of Gauss-Bonnet Holographic
Superconductors with Power-Maxwell Field}
\author[a,b]{Ahmad Sheykhi}
\author[a]{Hamid Reza Salahi}
\author[a]{Afshin Montakhab}
\affiliation[a]{ Physics Department and Biruni Observatory,
College of Sciences, Shiraz University, Shiraz 71454, Iran}
\affiliation[b]{Research Institute for Astronomy and Astrophysics
of Maragha (RIAAM), P.O. Box 55134-441, Maragha, Iran}
\emailAdd{asheykhi@shirazu.ac.ir}
\emailAdd{montakhab@shirazu.ac.ir}

\abstract{We provide an analytical as well as a numerical study of the
holographic $s$-wave superconductors in Gauss-Bonnet gravity with
Power-Maxwell electrodynamics. We limit our study to the case where
scalar and gauge fields do not have an effect on the background metric.
We use a variational method, based on Sturm-Liouville eigenvalue problem
for our analytical study, as well as a numerical shooting method in order
to compare with our analytical results. Interestingly enough, we observe
that unlike Born-Infeld-like nonlinear electrodynamics which decrease
the critical temperature compared to the linear Maxwell field, the
Power-Maxwell electrodynamics is able to increase the critical
temperature of the holographic superconductors in the sublinear
regime. We find that
requiring the finite value for the gauge field on the asymptotic
boundary $r\rightarrow \infty$, restricts the power parameter,
$q$, of the Power-Maxwell field  to be in the range $1/2<q<{(d-1)}/{2}$.
Our study indicates that it is quite possible to make condensation
\emph{easier} as $q$ \emph{decreases} in its allowed range.
We also find that for all values of $q$, the condensation can
be affected by the Gauss-Bonnet coefficient $\alpha$. However,
the presence of the Gauss-Bonnet term makes the transition
slightly harder. Finally, we obtain an analytic expression for
the order parameter and thus obtain the associated critical exponent near the
phase transition. We find that the critical exponent has its
universal value of $\beta=1/2$ regardless of the parameters $q$,
$\alpha$ as well as dimension $d$, consistent with mean-field values
obtained in previous studies.}
\begin{document}

\maketitle \flushbottom

%%%%%%%%%%%%%%%%%%%%%%%%%%%%%%%%%%%%%%%%%%%%%%%%%%%%%%%%%%%%%%%%
\section{Introduction}
It has been well established that the correspondence between
gravity in a $d$-dimensional anti-de Sitter (AdS) spacetime  and
conformal field theory (CFT) living on the $(d-1)$-dimensional
boundary of the corresponding spacetime, provides a powerful tool to study
strongly coupled systems \cite{Mal}. One of the intriguing
applications of gauge/gravity duality is to shed light on the
open problems in condensed-matter physics. Understanding the
mechanism governing the high-temperature superconductor systems
has been one of the most important challenges in condensed-matter
physics. The AdS/CFT correspondence provides a new approach for
calculating the properties of the superconductors using a dual
classical gravity description and may shed light on the pairing
mechanism in high temperature superconductors. Recently, it
has been shown that some properties of strongly
coupled superconductors can be potentially described by classical
general relativity living in one higher dimension, which is known
as \textit{holographic superconductors} \cite{Har}. It was suggested that
the instability of the bulk black hole corresponds to a second order
phase transition from normal state to superconducting state which
brings the spontaneous $U(1)$ symmetry breaking \cite{Gub}. The
first holographic $s$-wave superconductor model known as
Abelian-Higgs model was proposed by Hartnoll et al.,
\cite{Har,Har2}. This basically consists of a black hole and a
complex scalar field minimally coupled to an Abelian gauge field.
It was observed that a scalar hair is formed below a certain
critical temperature, $T_{c}$, due to the breaking of a local $U(1)$
gauge symmetry near the black hole event horizon.

Following this novel idea by Hartnoll et al., \cite{Har},
studies of holographic superconductors have attracted considerable
attention and become an active field of research. Consequently, various aspects
of the holographic superconductors, in the context of Einstein
gravity have been explored in the literature
\cite{Mus,RGC1,P.GWWY, P.BGRL, P.MRM, P.CW, P.ZGJZ,RGC2}. Generalization
to other gravity theories such as
Gauss-Bonnet gravity have also been considered. For example, holographic superconductors
with various condensates in Einstein-Gauss-Bonnet gravity were
explored in \cite{Wang1}. A general class of holographic
superconducting models with Gauss-Bonnet corrections terms was
studied in \cite{Wang2}. It was shown that different values of
Gauss-Bonnet correction term and model parameters can determine
the order of phase transitions and critical exponents of
second-order phase transition \cite{Wang2}. Also, using the variational
method for the Sturm-Liouville eigenvalue problem, some properties
of holographic superconductors with Gauss-Bonnet gravity in probe
limit were analytically calculated \cite{RGC3}. It was found that
the critical exponent of the condensation in Gauss-Bonnet gravity is
$\beta=1/2$ which is consistent with mean-field theory and previous
studies\cite{RGC3}. Other
studies on the holographic superconductors in the framework of
Gauss-Bonnet-AdS black holes were carried out in \cite{GB
HSC,Sub}.

Most of the works mentioned above have been done with the gauge
field as a linear Maxwell field. However, it is interesting to
investigate the effects of nonlinear electrodynamics on the
properties of the holographic superconductor both in Einstein as well as
Gauss-Bonnet gravities. Considering three types of nonlinear
electrodynamics, namely, Born-Infeld, Logarithmic and Exponential
nonlinear electrodynamics, and using numerical methods, it has
been observed that, in the Schwarzschild AdS black hole
background, the higher nonlinear electrodynamics corrections make
the condensation harder \cite{Zi}. The effects of an external
magnetic field on the holographic superconductors in the presence
of nonlinear corrections to the usual Maxwell action were examined
in \cite{DR}. Analytical study of holographic superconductors in
Born-Infeld electrodynamics have been carried out in
\cite{AnalyBI}. When the background metric is the Gauss-Bonnet AdS
black holes, several properties of holographic $s$-wave
superconductors with Born-Infeld electrodynamics were explored by
employing the Sturm-Liouville eigenvalue problem \cite{Gan1}.
Performing explicit analytic computations, Meissner-like effect in
the $(4+1)$-dimensional planar Gauss-Bonnet-AdS black hole
background with several nonlinear corrections to the gauge field
have also been studied \cite{Lala}.

Applying the Sturm-Liouville analytical and numerical methods
in the background of $d$-dimensional Schwarzschild AdS black hole,
the behavior of the holographic superconductors have been
explored by introducing a complex charged scalar field coupled
with a Power-Maxwell field, both in the probe limit \cite{PM} and
in the presence of backreaction \cite{PMb}. However, the
properties of the Gauss-Bonnet holographic superconductor from the
Power-Maxwell field have not been explored yet. In this paper, we
intend to extend the analytical study of holographic
superconductor, by taking into account both the higher order
Gauss-Bonnet curvature correction terms, as well as the nonlinear
Power-Maxwell electrodynamics. In particular, we shall disclose
the effects of these correction terms on the critical temperature
of the superconductor and its condensation. Interestingly, we find
that the effect of sublinear Power-Maxwell field can lead to the
easing of condensation formation and consequently a higher $T_c$.

This paper is outlined as follows: In the next section, we
introduce the basic field equations of holographic superconductors
in the background of Gauss-Bonnet AdS black holes when the gauge
field is in the form of Power-Maxwell field. In section \ref{Cri},
we use the Sturm-Liouville method and obtain a relation between
the critical temperature and charge density, and show that
$T_c$  increases for some values of Power-Maxwell parameter. In section
\ref{num}, we perform the numerical study and calculate the critical
temperature by applying the
shooting method. We also compare our analytical and numerical results in this section.
In section \ref{CriE}, we calculate the critical exponent and the
condensation values of the Power-Maxwell holographic superconductor and
show that it indeed has the universal value of $\beta= 1/2$. The last
section is devoted to concluding remarks.
%%%%%%%%%%%%%%%%%%%%%%%%%%%%%%%%%%%%%%%%%%%%%%%%%%%%%%%%%%%%%%

\section{Gauss-Bonnet holographic superconductors with Power-Maxwell field }
We consider the action of Einstein-Gauss-Bonnet-AdS gravity in
$d$-dimensions,
\begin{eqnarray}\label{einestein guass-bonnet action}
S_{G}=\int d^{d}x \sqrt{-g}\left[ ( R-2 \Lambda) +\frac{\alpha}{2}
(R^{2}-4 R^{\mu\nu}
R_{\mu\nu}+R^{\mu\nu\rho\sigma}R_{\mu\nu\rho\sigma}\right],
\end{eqnarray}
where $\alpha$ is the Gauss-Bonnet coefficient, $\Lambda
=-{(d-1)(d-2)}/{(2l^2)}$ is the cosmological constant of
$d$-dimensional AdS spacetime with radius $l$. For simplicity,
hereafter we set $l=1$. Since we consider the probe limit, the
gauge and the scalar fields do not back react on the background
metric. The line element of the $d$-dimensional spacetime with
flat horizon is written as \cite{Cai1}
\begin{eqnarray}\label{metric}
ds^{2}&=- f(r) dt^{2}+\frac{dr^2}{f(r)}+r^{2}dx_i dx^i,
\end{eqnarray}
where
\begin{eqnarray}\label{f(r)}
f(r)=\frac{r^2}{2 \alpha} \left[1-\sqrt{1- 4 \alpha \left(1-
\frac{r_{+}^{d-1}}{r^{d-1}}\right)}\right],
\end{eqnarray}
where $r_{+}$ is the positive real root, $f(r_{+})=0$. In the
asymptotic region where $r \rightarrow \infty$, we have
\begin{eqnarray}\label{fasym}
f(r) \approx \frac{r^2}{2 \alpha} (1-\sqrt{1-4\alpha}).
\end{eqnarray}
Here, the positivity of $f(r)$ implies $0\leq\alpha\leq1/4$ and real. The
Hawking temperature of black hole on the horizon $r_{+}$, which
will be interpreted as the temperature of the CFT, may be obtained
as \cite{Wang1}
\begin{eqnarray}\label{T}
T=\frac{f'(r_{+})}{4 \pi}=\frac{(d-1) r_{+}}{4\pi},
\end{eqnarray}
where prime denotes derivative with respect to $r$. If we define
the effective AdS radius as
 \begin{eqnarray}\label{Leff}
L_{\rm eff}^2=\frac{2 \alpha}{1 - \sqrt{1 - 4 \alpha}},
\end{eqnarray}
then, in the asymptotic limit where $r\rightarrow
\infty$, $f(r)$ can be written as
\begin{eqnarray}\label{finfty}
 f(r) = \frac{r^2}{L_{\rm eff}^2},
 \end{eqnarray}
which has the form of Schwarzschild AdS black hole for
$r\rightarrow\infty$. Now, consider the action of matter which is
made of complex charged scalar field in bulk, that minimally
couples to the Power-Maxwell field:
 \begin{eqnarray}\label{matteraction}
 S_{m}=\int d^{d}x \sqrt{-g}\left[- b (F_{\mu\nu}F^{\mu\nu})^q  - |\nabla\psi- i  A \psi|^2 - m^2 |\psi|^2 \right
 ],
\end{eqnarray}
where $F^{\mu\nu}$ is the electromagnetic field tensor, $A$  and
$\psi$ are respectively, the gauge and complex charged scalar
field. Here, $b$ is a constant and $q$ is the power parameter
of the Power-Maxwell field. In the case where $b \rightarrow
1/4$ and $q \rightarrow1$, the Power-Maxwell Lagrangian will reduce to the
Maxwell case. Throughout this work, we set the charge of the
scalar field $\psi$ equal to unity. Varying the matter action
(\ref{matteraction}) with respect to the scalar and gauge fields
yield the following field equations,
\begin{eqnarray}
(\nabla_{\mu}-iA_{\mu})(\nabla^{\mu}-iA^{\mu})\psi-m^2\psi=0,
\end{eqnarray}
\begin{eqnarray}
4 b q\nabla_{\nu}\left[(F_{\sigma \delta}F^{\sigma
\delta})^{q-1} F^{\mu\nu}\right]-2A^{\mu} |\psi|^2+i(\psi
\nabla^{\mu}\psi^*-\psi^*\nabla^{\mu}\psi)=0.
\end{eqnarray}
We shall assume $\psi$ and $\phi$ to be real and only function of
$r$, i.e., $\psi$=$\psi$($r$) and $A=\phi(r)dt$. Thus, the
equations of motion for the gauge and scalar fields can be written as,
\begin{eqnarray}\label{eomphi}
\phi''+ \left(\frac{d-2}{2q-1}\right) \frac{\phi'}{r}+ \frac{\phi
\psi^2 \phi'^{2-2q}}{(-2)^{q} b q (2q-1)f}=0,
\end{eqnarray}
\begin{eqnarray}\label{eompsi}
\psi''+ \left(\frac{f'}{f}+\frac{d-2}{r}\right)\psi'+
\left(\frac{\phi^2}{f^2}-\frac{m^2}{f}\right)\psi=0.
\end{eqnarray}
Imposing the regularity conditions at the horizon $r_{+}$, yield
\begin{eqnarray}\label{regularity condition}
\phi (r_{+})=0 ,   \     \  \   \   \  \ \  \   \   f'(r_{+})
\psi'(r_{+})=m^2 \psi(r_{+}).
\end{eqnarray}
Now, we solve Eqs.~(\ref{eomphi}) and (\ref{eompsi}) in the
asymptotic region ($r\rightarrow \infty$). We find
\begin{eqnarray}\label{BC1}
&& \phi \approx \mu -
\frac{\rho^{\frac{1}{2q-1}}}{r^{\frac{d-2}{2q-1} - 1}},\\
&& \psi \approx \frac{\psi^{-}}{r^{\lambda_{-}}}
+\frac{\psi^{+}}{r^{\lambda_{+}}}, \label{BC2}
\end{eqnarray}
where $\mu$ and $\rho$ are respectively interpreted as the
chemical potential and charge density in the dual field theory,
and
\begin{eqnarray}\label{lambda}
\lambda_{\pm} = \frac{1}{2}\left[(d-1) \pm  \sqrt {(d-1)^2 + 4m^2
L_{\rm eff}^2}\right].
\end{eqnarray}
According to AdS/CFT correspondence, $\psi_{+}$ and $\psi_{-}$ are
normalizable modes of the scalar field $\psi$ which can be
regarded as the source of the dual operator $\mathcal{O}$,
$\psi_{-}=<\mathcal{O_{-}}>$ and $\psi_{+}=<\mathcal{O_{+}}>$,
respectively. Following \cite{Har,Har2}, we can impose the
boundary condition in which either $\psi_{+}$ or $\psi_{-}$
vanishes, so that the theory is stable in the asymptotic AdS
region. It is worth noting that the gauge field $\phi$  depends on
the power parameter $q$ of the Power-Maxwell field at the
asymptotic AdS region ($r\rightarrow \infty$). This is in contrast
to the case of holographic superconductor with other nonlinear
electrodynamics such as Born-Infeld \cite{AnalyBI} and
Born-Infeld-like electrodynamics \cite{Zi}.

At the boundary where $r\rightarrow \infty$, the gauge field
should have a finite value. Thus Eq.~(\ref{BC1}) implies that
$\frac{d-2}{2q-1}-1>0$, which restricts the values of $q$ to be
$q<{(d-1)}/{2}$. On the other hand since $\frac{d-2}{2q-1}>1$,
it must then be a positive real number, which considering that
$d-2>0$, implies that  $2q-1>0$ or $q>1/2$.  We have therefore
simply extracted the meaningful range of the parameter $q$ to be
$1/2<q<{(d-1)}/{2}$.

%%%%%%%%%%%%%%%%%%%%%%%%%%%%%%%%%%%%%%%%%%%%%%%%%%%%%%%%%%

\section{Critical temperature in terms of charge density} \label{Cri}
In this section, we obtain a relation between the
critical temperature and charge density  of the holographic
superconductor in Gauss-Bonnet-AdS black holes in the presence
of Power-Maxwell gauge field. We shall use the Sturm-Liouville
eigenvalue problem and limit our study to the case where the
scalar and gauge field do not affect on the background metric.

For this purpose, we first transform the coordinate $r$ to $z$,
where $z={r_{+}}/{r}$. In this new coordinate,
the equations of motion (\ref{eomphi}) and (\ref{eompsi}) are
rewritten as,
\begin{eqnarray}\label{eomphiz}
\phi'' + \left(\frac{4q-d}{2q-1}\right)\frac{\phi'}{z} +
\frac{\phi \psi^2 \phi'^{2-2q}r_{+}^{2q}}{2^q (-1)^{3q} b q
(2q-1) z^{4q} f}=0
\end{eqnarray}
\begin{eqnarray}\label{eompsiz}
\psi'' + \left(\frac{f'}{f}-\frac{d-4}{z}\right)\psi' +
\frac{r_{+}^2}{z^4}\left(\frac{\phi^2}{f^2}-
\frac{m^2}{f}\right)\psi=0,
\end{eqnarray}
where the prime now indicates the derivative with respect to the
new coordinate $z$. Very close the critical temperature $T_{c}$,
we have $\psi\rightarrow 0$, which implies that the condensation
approaches zero. Therefore, Eq.~(\ref{eomphiz}) reduces to
\begin{eqnarray}\label{phipsizero}
\phi'' + \left(\frac{4q-d}{2q-1}\right)\frac{\phi'}{z} =0.
\end{eqnarray}
Using the boundary conditions (\ref{regularity condition}) and
(\ref{BC1}), one can easily show that, near the critical
temperature ($r_{+}=r_{+c}$), Eq.(\ref{phipsizero}) has a
solution of the form
\begin{eqnarray}\label{phisolution}
\phi=\zeta r_{+c} \left(1-z^{\frac{d-2}{2q-1}-1}\right),
\end{eqnarray}
where $\zeta= (\frac{\rho}{r_{+}^{d-2}})^{\frac{1}{2q-1}}$. Next,
we introduce a trial function $F(z)$ as in \cite{Sio} and rewrite the
scalar field $\psi$, near the boundary,
\begin{eqnarray}\label{psibound}
\psi\mid_{z=0}  \approx \frac{\psi_{i}}{r^{\lambda_i}}=
<\mathcal{O}_{i}> \frac{z^{\lambda_i}}{r_{+}^{\lambda_i}} F(z),
\end{eqnarray}
where $i=(+, -)$. The trial function near the boundary $z=0$
satisfies the boundary conditions $F(0)=1$ and $F'(0)=0$
\cite{Sio}. Substituting Eqs.~(\ref{f(r)}), (\ref{phisolution})
and (\ref{psibound}), in  Eq.~(\ref{eompsiz}), one obtains
\begin{align}\label{F}
&&F'' (z) +\Bigg{\{} \frac{2\lambda_{i}}{z}+\frac{-2(d-3) z^{d-1}
\alpha +(d-2) (-1+4\alpha+\sqrt{1+4(z^{d-1}-1)\alpha})} {z
\sqrt{1+4(z^{d-1}-1)\alpha}\left(-1+\sqrt{1+4(z^{d-1}-1)\alpha}\right)}\Bigg{\}}F'
(z)\nonumber
\\&&+
\Bigg{\{}\frac{4(-1+z^{\frac{d-2q-1}{2q-1}})\alpha^2\zeta^2}{(-1+\sqrt{1+4(z^4-1)\alpha})^2}
+\frac{\lambda_{i}
(\lambda_{i}-1)}{z^2}+\frac{2m^2\alpha}{z^2(-1+\sqrt{1+4(z^{d-1}-1)})}\nonumber
\\ &&-\frac{(2(d-3)z^{d-1}
\alpha+(d-2)(-1+\sqrt{1+4(z^{d-1}-1)\alpha})\lambda_{i}}{z^2\sqrt{1+4(z^{d-1}-1)\alpha}
(-1+\sqrt{1+4(z^{d-1}-1)\alpha})}\Bigg{\}}
F(z)=0.
\end{align}
The above equation is a second order ordinary differential
equation in the form of
\begin{eqnarray}
F''(z) + p(z)F'(z) +q(z)F(z)+\zeta^2 w(z)F(z)=0,
\end{eqnarray}
where
\begin{eqnarray}
p(z)= \frac{2\lambda_{i}}{z}+\frac{-2(d-3) z^{d-1}\alpha +(d-2)
(-1+4\alpha+\sqrt{1+4(z^{d-1}-1)\alpha})}{z
\sqrt{1+4(z^{d-1}-1)\alpha}(-1+\sqrt{1+4(z^{d-1}-1)\alpha})},
\end{eqnarray}
\begin{align}
q(z)=&\frac{\lambda_{i} (\lambda_{i}-1)}{z^2}-
\frac{(2(d-3)z^{d-1}
\alpha+(d-2)(-1+\sqrt{1+4(z^{d-1}-1)\alpha})\lambda_{i}}{z^2\sqrt{1+4(z^{d-1}-1)\alpha}(-1+\sqrt{1+4(z^{d-1}-1)\alpha})}\nonumber
\\&+\frac{2m^2\alpha}{z^2[-1+\sqrt{1+4(z^{d-1}-1)}]},
\end{align}
\begin{eqnarray}
w(z)=\frac{4(-1+z^{\frac{d-2q-1}{2q-1}})\alpha^2\zeta^2}{[-1+\sqrt{1+4(z^4-1)\alpha}]^2}.
\end{eqnarray}
It is a matter of calculations to convert Eq.~(\ref{F}) to the
standard form of the Sturm-Liouville equation,
\begin{eqnarray}\label{SL}
[T(z)F'(z)]' - Q(z) F(z)+\zeta^2 P(z) F(z)=0,
\end{eqnarray}
where
\begin{eqnarray}\label{Tz}
&& Q(z)= -T(z)q(z), \   \   \   P(z)=T(z)w(z),\\
&& T(z)= z^{2 \lambda +2-d}\sqrt{1-z^{d-1}}
\exp\Bigg\{-\frac{z^{d-1}
\tilde{F_1}\left(1,\frac{1}{2},1,2;\frac{4 z^{d-1} \alpha }{4
\alpha -1},z^{d-1}\right)}{2 \sqrt{1-4 \alpha }}\Bigg\}.
\end{eqnarray}
Here $\tilde{F_1}$ is the Appell hypergeometric function \cite{PApp} with
two variables $\alpha$ and $z$. Expanding Eq.~(\ref{Tz}) for small
$\alpha$ and keeping terms up to
$\mathcal{O}\left(\alpha^2\right)$, we get
\begin{eqnarray}
&&T(z)\approx z^{2 \lambda +2-d} \left(z^{d-1}-1\right) \Bigg\{
1-\alpha  z^{d-1}+\alpha ^2 \left(2 z^{2 d-2}-3
z^{d-1}\right)+\mathcal{O}\left(\alpha^3\right)\Bigg\},
\end{eqnarray}

\begin{center}
\begin{table}[ht]
\begin{tabular}{|c|c|c|c|c|c|c|c|c|}
\hline
&\multicolumn{2}{c|}{$\alpha=0.05$}
&\multicolumn{2}{c|} {$ \alpha=0.1$}
&\multicolumn{2}{c|}{$\alpha=0.15$}
&\multicolumn{2}{c|}{$\alpha=0.2$} \\
\hline
$q$ &$a$ & $\zeta_{min}^2$ & $a $ &$\zeta_{min}^2$ & $a$ & $\zeta_{min}^2$ &$a$ & $\zeta_{min}^2$ \\
\hline
$5/4$    &0.7644&47.7162 &0.7579 & 53.1425 & 0.7488&60.3734&0.7358&70.5134\\
\hline
$1$  &0.7146&19.9456&0.7050&22.1278&0.6917&25.0053&0.6730&28.9837\\
\hline
$3/4$  &0.5878&8.6847&0.5693&9.5348&0.5442&10.6247&0.5092&12.0771\\
\hline
\end{tabular}
\caption{Analytical results for $\zeta^2_{\rm min}$ and $a$
with different values of the power parameter $q$ and Gauss-Bonnet
parameter $\alpha$ for $\lambda_+$. Here we have taken $d=5$ and
$\tilde{m}^2=-3$.}\label{tab1}
\end{table}
\end{center}

\begin{eqnarray}
Q(z) \approx -z^{2 \lambda -d}&&\Bigg\{\lambda  \Big[(d-\lambda
-1)+2 \alpha ^2 (2 d+\lambda -2) z^{3 d-3}+\left(3 \alpha
^2+\alpha +1\right) \lambda  z^{d-1} \nonumber
\\&& -\alpha  (5 \alpha +1) (d+\lambda -1) z^{2 d-2}\Big]+\tilde{m}^2 \Bigg\},
\end{eqnarray}
\begin{eqnarray}
P(z)\approx -\frac{z^{\frac{d (3-2 q)}{2 q-1}+2 \lambda
}}{z^{d-1}-1} \left(z^{\frac{d}{1-2 q}}-z^{\frac{2 q+1}{1-2
q}}\right)^2 && \Bigg\{\alpha ^2 z^{2 d}-
\nonumber
\\ && \alpha  (\alpha +1) z^{d+1}
+[\alpha (\alpha +2)-1] z^2 \Bigg\}.
\end{eqnarray}

\begin{figure}
\centering
\begin{subfigure}{.5\textwidth}
  \centering
  \includegraphics[width=1\linewidth]{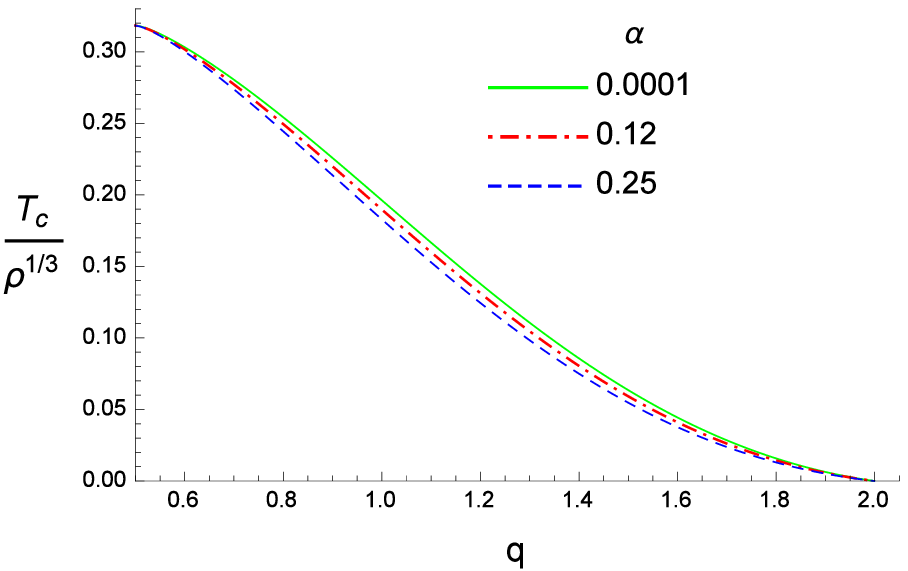}
  \caption{$\tilde{m}^2=-3$}
  \label{fig:sub1}
\end{subfigure}%
\begin{subfigure}{.5\textwidth}
  \centering
  \includegraphics[width=1\linewidth]{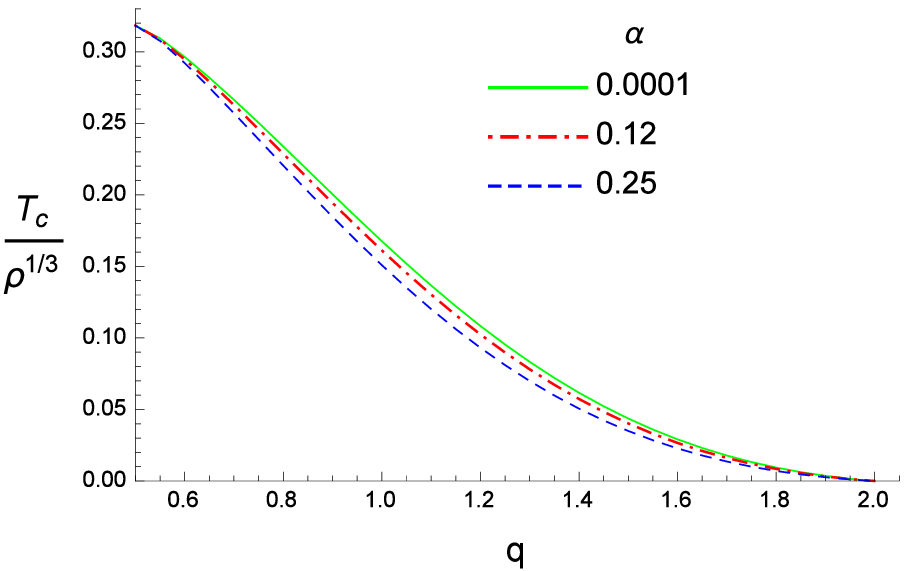}
  \caption{$\tilde{m}^2=0$}
  \label{fig:sub2}
\end{subfigure}
\caption{Rescaled critical temperature (${T_c}/{\rho^{1/3}}$) as a function of $q$ for
$d=5$.}
\label{fig1}
\end{figure}

We have also taken $\tilde{m}^2= m^2 L_{\rm eff}^2$ where $L_{\rm
eff}$ is given by Eq.~(\ref{Leff}). According to the Sturm-Liouville
eigenvalue problem, the eigenvalues of Eq.~(\ref{SL}) are obtained
as
\begin{eqnarray}\label{zeta2}
\zeta^2=\frac{\int_{0}^{1}[T(z)[F'(z)]^2+Q(z)F^2(z)]dz}{\int_{0}^{1}T(z)F^2(z)dz}.
\end{eqnarray}
Our strategy is to obtain the minimum  value of $\zeta^2$ by
varying Eq.~(\ref{zeta2}). Let us choose the trial function as
$F(z)=1-az^2$ which satisfies the boundary conditions. In the
remaining part of this section we shall concentrate on
$\lambda_{+}$.

Calculating the integrals in Eq.~(\ref{zeta2}) for $d=5$, $q=1$,
$\tilde{m}^2=-3$, and $\alpha=0.1$ we get
\begin{eqnarray}\label{zeta}
\zeta^2=\frac{128.654 a-199.113}{a (a-3.4531)+3.6591}+85.0957,
\end{eqnarray}
which has a minimum $\zeta^2_{\rm min}=22.1278$ for $a=0.7050$. By
choosing $d=5$, $q=3/4$, $\tilde{m}^2=-3$ and $\alpha=0.2$, for
$\zeta^2$ we arrive at
\begin{eqnarray}
 \zeta^2=\frac{28.096 a-33.1738}{a (1. a-3.04098)+2.64731}+25.9692.
\end{eqnarray}
The minimum will now be $\zeta^2_{\rm min}=12.0771$ at
$a=0.5092$. We summarize our results for $\zeta^2_{\rm min}$ in
Table \ref{tab1}.\par{}
Next, we calculate the critical temperature.  At the critical point
we have $r_{+}=r_{+c}$, Eq.~(\ref{T}) leads to
\begin{eqnarray}
T_c=\frac{(d-1) r_{+c}}{4\pi}.
\end{eqnarray}
Using the fact that $ \zeta_{\rm
min}=\left(\frac{\rho}{r_{+c}^{d-2}}\right)^{\frac{1}{2q-1}}$, we
finally reach
\begin{eqnarray}\label{Tc}
T_{c}= \gamma \rho^{\frac{1}{d-2}}, \ \ \ \ \ \ \ \ \ \
 \gamma=\frac{(d-1)}{4\pi} \zeta_{\rm min}^{\frac{1-2q}{d-2}}.
\end{eqnarray}

\begin{figure}
\centering
\begin{subfigure}{.5\textwidth}
  \centering
  \includegraphics[width=1\linewidth]{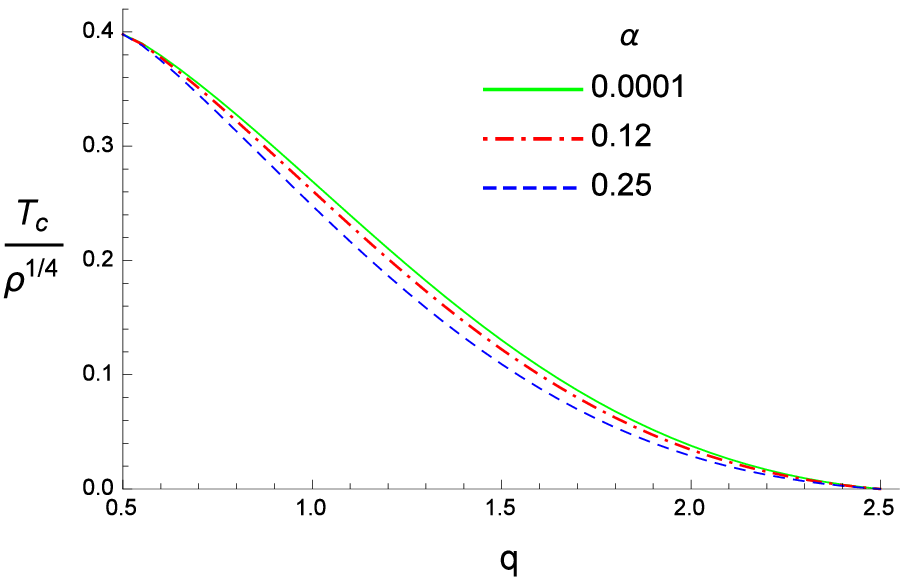}
  \caption{$\tilde{m}^2=-4$}
  \label{fig:sub1}
\end{subfigure}%
\begin{subfigure}{.5\textwidth}
  \centering
  \includegraphics[width=1\linewidth]{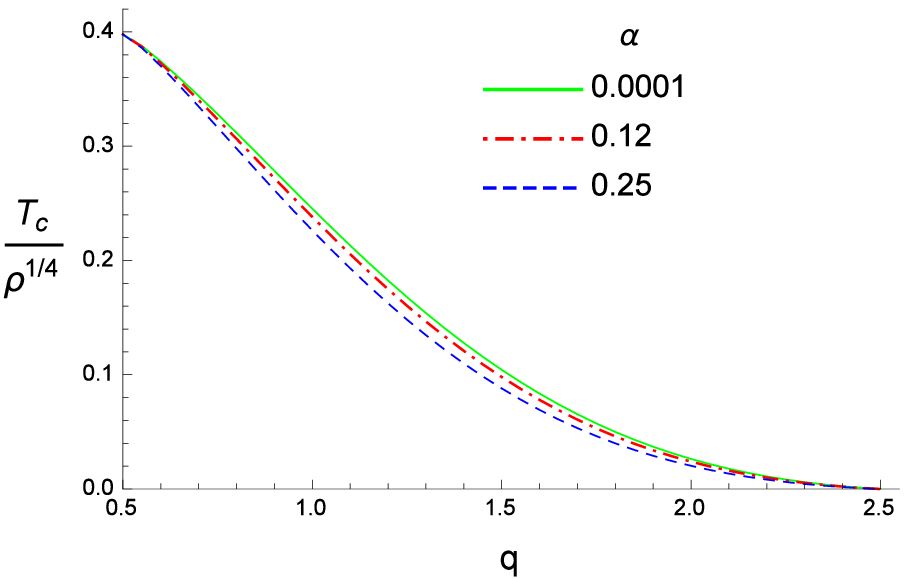}
  \caption{$\tilde{m}^2=0$}
  \label{fig:sub2}
\end{subfigure}
\caption{Rescaled critical temperature (${T_c}/{\rho^{1/4}}$) as a function of $q$ for $d=6$.}
\label{fig2}
\end{figure}

We now provide a systematic presentation of the critical
temperature as a function of various variables and parameters.
The critical temperature can change continuously as a function of
$q$ in the allowed range.  Therefore, in Fig.~(\ref{fig1}) we
provide such a variation for three different values of $\alpha$
and two different values of $\tilde{m}^2$ for $d=5$. Clearly,
increasing $\alpha$ decreases the critical temperature. However,
and more importantly, the critical temperature is a decreasing
function of $q$, indicating the possibility of easing the
superconducting phase transition with sublinear Power-Maxwell
field. It is also interesting to note that the behavior of the
critical temperature at both lower and upper bound of $q$. In the
lower bound, where the maximum occurs, the results are independent
of both $\alpha$ as well as $\tilde{m}^2$. This can clearly be
seen from Eq.~(\ref{Tc}) where $\gamma$ becomes only a function of
$d$ in this limit. On the other hand, the superconducting phase
transition vanishes as $q$ approaches its upper bound. This can
also be seen from Eq.~(\ref{phisolution}) where the gauge field
vanishes for $q\rightarrow(d-1)/2$, where $\psi=0$ becomes the
only solution. This is consistent with the understanding that the
electrostatic repulsion must overcome the gravitational attraction
for the formation of the scalar field \cite{GuPu}. We also provide
the same information in one higher dimensions, $d=6$, in
Fig.~(\ref{fig2}), where the critical temperature now scales with
$\rho^{1/4}$. Here, the general trend discussed above is the same
except that the range of possible $q$ values has increased and
that the transition point at the lower bound ($q=1/2$) increases
with increasing $d$. This is also consistent with the general
understanding that phase transitions are easier to achieve in
higher dimensional systems.

\begin{figure}
\centering
\begin{subfigure}{.5\textwidth}
  \centering
  \includegraphics[width=1\linewidth]{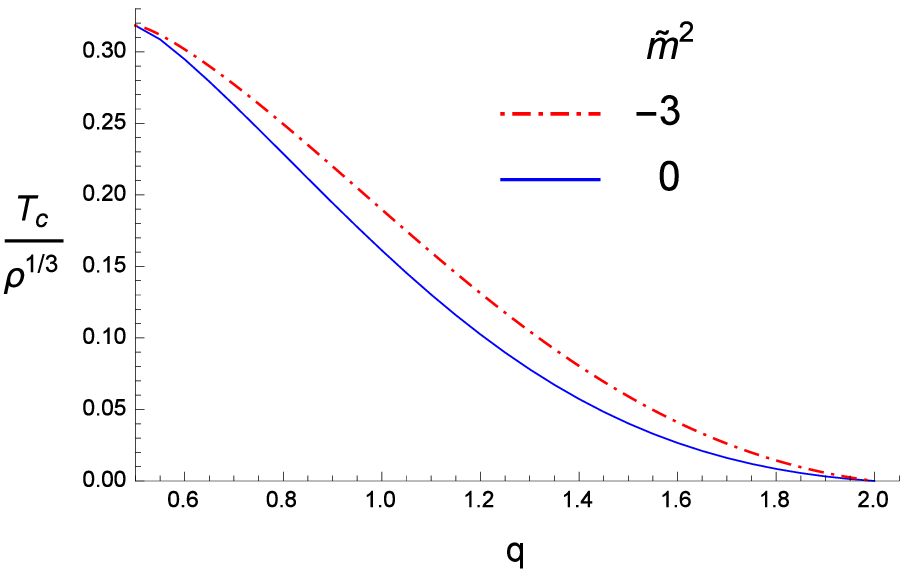}
  \caption{d=5}
  \label{fig:sub1}
\end{subfigure}%
\begin{subfigure}{.5\textwidth}
  \centering
  \includegraphics[width=1\linewidth]{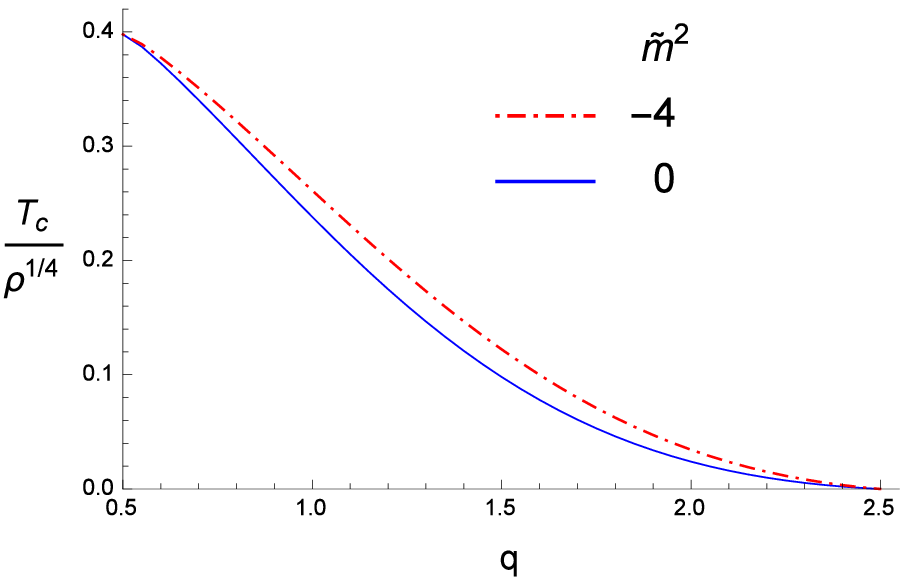}
  \caption{$d=6$}
  \label{fig:sub2}
\end{subfigure}
\caption{Rescaled critical temperature (${T_c}/{\rho^{1/(d-2)}}$) as a function of $q$
for a fixed value of $\alpha =0.12$ }
\label{fig3}
\end{figure}

\begin{figure}
\centering
  \includegraphics[width=.7\linewidth]{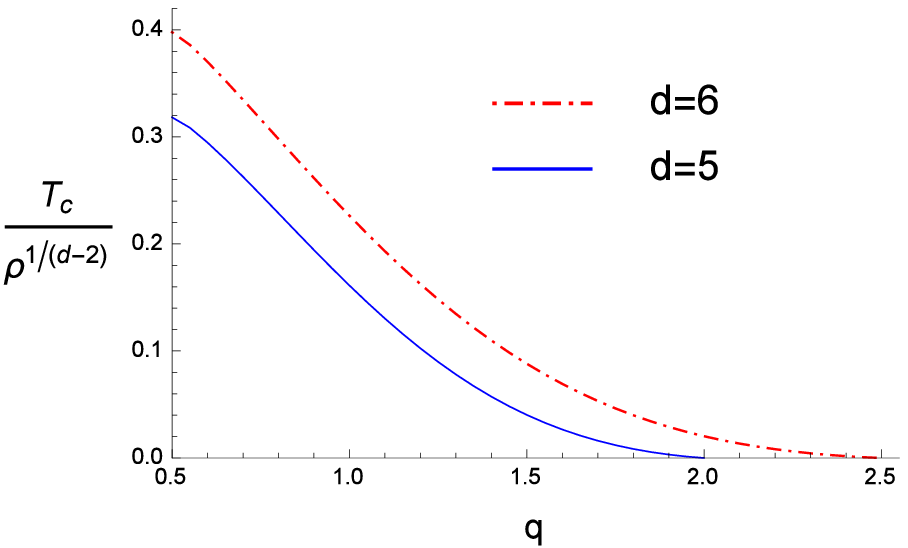}
  \caption{Rescaled critical temperature (${T_c}/{\rho^{1/(d-2)}}$) as a function of $q$
  for a fixed value of $\tilde{m}^2=0$ and $\alpha=0.12$}
\label{fig4}

\end{figure}

In order to more explicitly show the effect of $\tilde{m}^2$,
Fig.~(\ref{fig3}) provides
such a plot for (a) $d=5$ and (b) $d=6$, where one can clearly see that
decreasing
$\tilde{m}^2$ increases the critical temperature for a given value of
$q$.  Furthermore,
Fig.~(\ref{fig4}) shows the easing of superconducting phase transition in
a higher dimensional system.

%%%%%%%%%%%%%%%%%%%%%%%%%%%%%%%%%%%%%%%%%%%%%%%%%%%%%%%%%%%%%%%%%%%%%%%%%%%%%%

\begin{center}
\begin{table}[ht]
\begin{tabular}{|c|c|c|c|c|c|c|c|}
\hline
&\multicolumn{2}{c|}{$\alpha=0.05$}
&\multicolumn{2}{c|} {$ \alpha=0.1$}
&\multicolumn{2}{c|}{$\alpha=0.15$} \\
\hline
$q$ & Analytical & Numerical &Analytical & Numerical
&Analytical & Numerical  \\
\hline
$5/4$   & 0.1211 $\rho^{1/3}$ & 0.1248 $\rho^{1/3}$ & 0.1179
$\rho^{1/3}$ & 0.1212 $\rho^{1/3}$ & 0.1142 $\rho^{1/3}$&0.1169
$\rho^{1/3}$ \\
\hline
$1$  &0.1933 $\rho^{1/3}$&0.1949 $\rho^{1/3}$&0.1900 $\rho^{1/3}$&0.1914
$\rho^{1/3}$&0.1861 $\rho^{1/3}$&0.1871 $\rho^{1/3}$ \\
\hline
$3/4$  &0.2658 $\rho^{1/3}$&0.2659 $\rho^{1/3}$&0.2638
$\rho^{1/3}$&0.2638 $\rho^{1/3}$&0.2614 $\rho^{1/3}$&0.2614 $\rho^{1/3}$\\
\hline
\end{tabular}
\caption{Comparison of analytical and numerical values of critical temperature
for $d=5$ and $\tilde{m}^2=-3$ for various values of $q$ and $\alpha$.}\label{tab2}
\end{table}
\end{center}
\section{Critical Temperature from numerical method} \label{num}
In this section, we propose to numerically study the critical
behavior of the Power-Maxwell holographic superconductor in
Gauss-Bonnet gravity. For this purpose, we adopt the shooting
method \cite{}. Let us note that Eqs. (\ref{eomphi}) and (\ref{eompsi})
remain unchanged under transformation
\begin{eqnarray}\label{scale}
r \rightarrow ar, \ \ \ \ \ f \rightarrow a^2f, \ \ \ \ \ \phi \rightarrow a\phi, \ \ \ \ \ \psi \rightarrow \psi .
\end{eqnarray}
Taking advantages of this scaling symmetry, we can set $r_{+}$ as
unity in our numerical calculations and we can build dimensionless
quantities such as $T/\rho^{\frac{1}{d-2}}$.

\begin{figure}
  \begin{subfigure}{0.5\linewidth}
    \centering
    \includegraphics[width=1\linewidth]{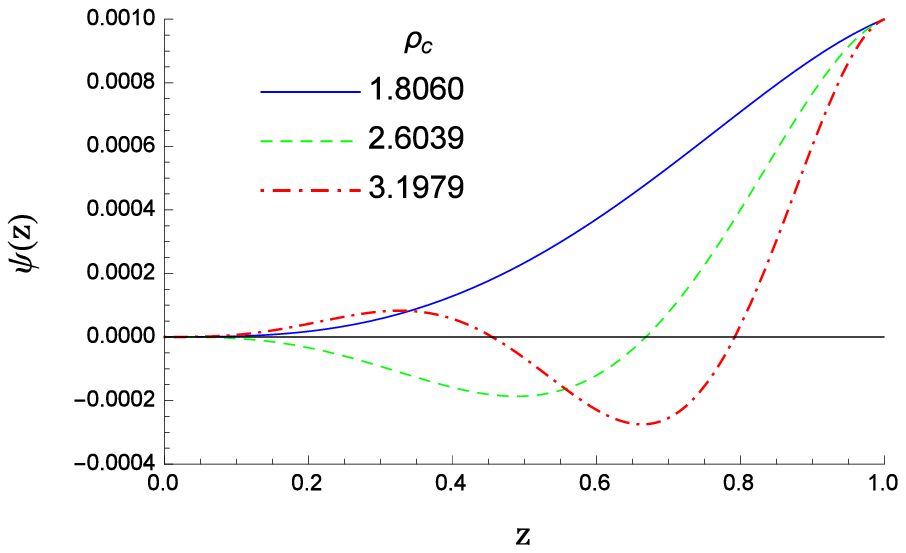}
    \caption{$q=3/4$, $\alpha=0.05$, $\tilde{m}^2=-3$}
    \label{fig5a}
    \vspace{4ex}
  \end{subfigure}%%
  \begin{subfigure}{0.5\linewidth}
    \centering
    \includegraphics[width=1\linewidth]{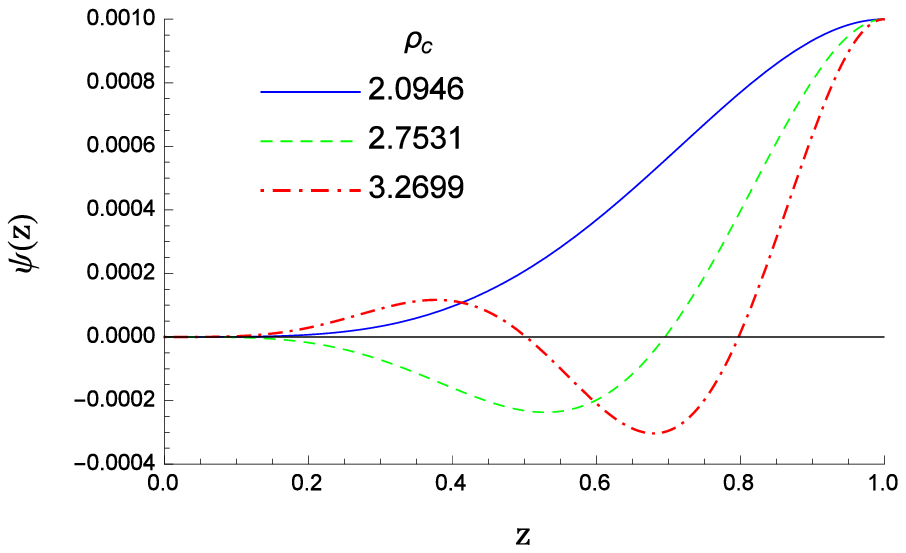}
    \caption{$q=3/4, \alpha=0.15, \tilde{m}^2=0$}
    \label{fig5b}
    \vspace{4ex}
  \end{subfigure}
  \begin{subfigure}{0.5\linewidth}
    \centering
    \includegraphics[width=1\linewidth]{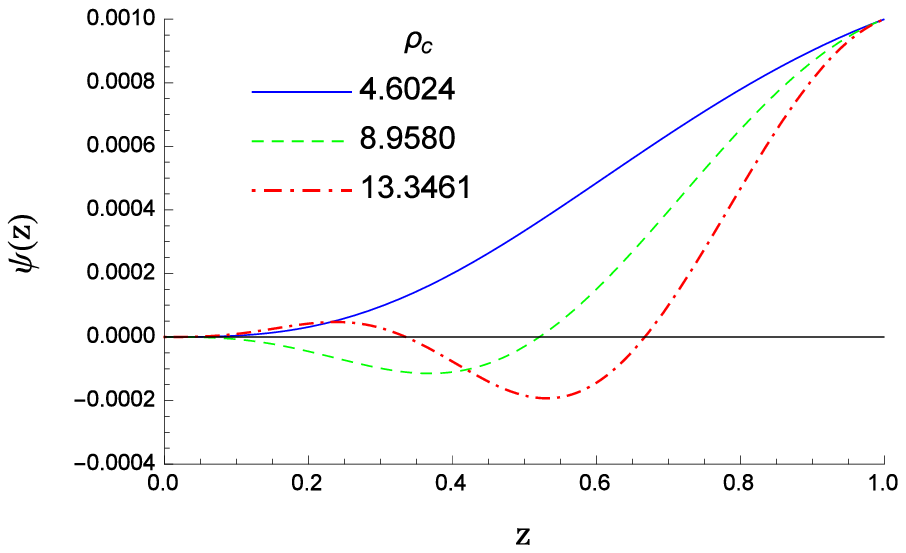}
    \caption{$q=1, \alpha=0.1, \tilde{m}^2=-3$}
    \label{fig5c}
  \end{subfigure}%%
  \begin{subfigure}{0.5\linewidth}
    \centering
    \includegraphics[width=1\linewidth]{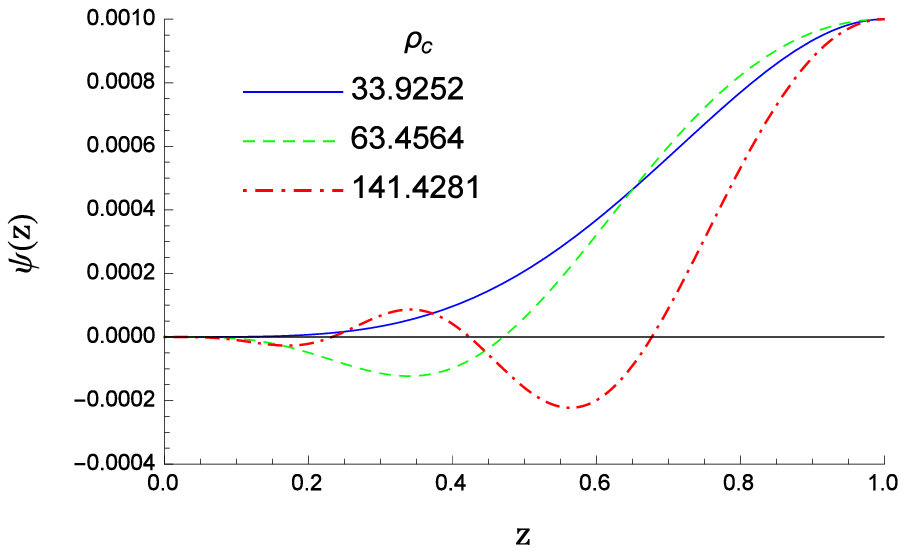}
    \caption{$q=5/4, \alpha=0.05, \tilde{m}^2=0$}
    \label{fig5d}
  \end{subfigure}
  \caption{$\psi(z)$ for some values of $q$, $\alpha$ and $\tilde{m}^2$
  in five dimensions with boundary condition $\psi^-=0$ for three lowest
  values of critical charge density. In each of these graphs, blue line
  shows the lowest-lying $\rho_c$.}
  \label{fig5}
\end{figure}

Because we want to obtain critical temperature, we will take
$\phi$ as (\ref{phisolution}). Thus the equation of motion for
$\psi$ in the $z$ coordinate becomes
\begin{eqnarray}
\psi'' + \left(\frac{f'}{f}-\frac{d-4}{z}\right)\psi' +
\frac{1}{z^4}\left(\frac{\rho ^{\frac{2}{2 q-1}} \left(1-z^{\frac{d-2}{2 q-1}-1}\right)^2}{f^2}-
\frac{m^2}{f}\right)\psi=0.
\end{eqnarray}
The behavior of $\psi$ near the horizon boundary is
\begin{eqnarray}
\psi \approx \psi(1) + \psi'(1) (1-z) + \frac{\psi''(1)}{2} (1-z)^2+\dotsc,
\end{eqnarray}
and near the infinite boundary is
\begin{eqnarray}
\psi \approx \psi^- z^{\lambda_-} + \psi^+ z^{\lambda_+}.
\end{eqnarray}
Using the equation of motion for $\psi$, we can find $\psi'(1)$ and
$\psi''(1)$ in the terms of $\psi(1)$ which is an arbitrary
constant. Considering the fact that we are close to the critical
point and $\psi$ is very small, we choose $\psi(1)=0.001$.
According to the shooting method, for specific dimension $d$ and
reduced scalar field mass $\tilde{m}^2$, we can perform numerical
calculation near the horizon boundary with one shooting parameter
$\rho$ to get proper solutions at the infinite boundary. For
specific values of $\rho$, boundary condition $\psi_{-}=0$ (or
$\psi_{+}=0$) is satisfied. In Fig. (\ref{fig5}), we plot
$\psi(z)$ for three first critical charge densities for some
values of $q$, $\tilde{m}^2$ and $\alpha$ in five dimensions. It is
believed that the first condensation occurs for the lowest value of
$\rho_c$ and condensations due to higher values of $\rho_c$ are not
considered to be stable \cite{GuPu}. Having the lowest $\rho_c$ and using
dimensionless quantity $T/\rho^{\frac{1}{d-2}}$, we easily obtain the
critical temperature for different values of $q$ and $\alpha$. In Table
\ref{tab2} we compare the values of the critical temperature from
the numerical results based on shooting method with boundary
condition $\psi_{-}=0$ for $\tilde{m}^2=-3$ with the analytical
results obtained in the previous section. We see that our
analytical results obtained by Sturm-Liouville method are in good
agreement with the numerical results from shooting method.

\section{Critical exponent} \label{CriE}
In this section, we propose to analytically calculate the critical
exponent of the Gauss-Bonnet holographic superconductor in the
presence of the Power-Maxwell field. Near the critical
temperature, $T\approx T_c$, the condensation value
$<\mathcal{O}_{i}>$ is very small. Inserting Eq.~(\ref{psibound})
into Eq.~(\ref{eomphiz}), we get
\begin{eqnarray}\label{phipsismall}
\phi'' +\left(\frac{4q-d}{2q-1}\right) \frac{1}{z} \phi'
+\frac{r_{+}^{2q-2\lambda_{i}-2} z^{2\lambda_{i}-4q}F^2
\phi'^{2-2q} \phi <\mathcal{O}_i>^2}{(-1)^{3q} 2^q b q (2q-1)
g(z)}=0,
\end{eqnarray}
where $g(z)=\frac{1-\sqrt{1-4 \alpha (1-z^{d-1})}}{2 \alpha z^2}$.
One can easily see that for $<\mathcal{O}_i>=0$, Eq.~(\ref{phipsismall})
has a solution of the form
(\ref{phisolution}). Because $<\mathcal{O}_i>$ is very small, we
assume that solution of Eq.~(\ref{phipsismall}) has the form
\begin{eqnarray}\label{phisolsmallpsi}
\phi(z)= AT_{c}(1-z^{\frac{d-2}{2q-1}-1})-(A T_{c})^{p}\left[\frac
{r_{+}^{2q-2\lambda_{i}-2}<\mathcal{O}_{i}>^2}{(-1)^{3q}2^qb
q(2q-1)} \right]\chi (z),
\end{eqnarray}
where $A=\frac{4\pi}{d-1}\zeta_{min}$. Substituting Eq.~(\ref{phisolsmallpsi}) into Eq.~(\ref{phipsismall}) and keeping terms
up to the linear order in $\alpha$, we get
\begin{eqnarray}\label{chi}
\chi''+\left(\frac{4q-d}{2q-1}\right)\frac{\chi'}{z}-
\frac{(1-\frac{d-2}{2q-1})^{2-2q}z^{\eta}(1-z^{\frac{d-2}{2q-1}-1})F(z)^2}{(z^4-1)[(z^4-1)\alpha-1]
}=0,
\end{eqnarray}
where
\begin{eqnarray}
\eta=2\lambda-4q+\left(\frac{d-2}{2q-1}-2\right)(2-2q)+2.
\end{eqnarray}
and we have set $p=3-2q$. Note that this choice for $p$ allows us
to obtain an equation for $\chi$ independent of $T_c$ and $A$.
Near the boundary, the solution of $\phi$ given in Eq.~(\ref{BC1}),
after using $r={r_+}/{z}$, can be written as
\begin{eqnarray}\label{phiz}
\phi(z) = \mu -
\frac{\rho^{\frac{1}{2q-1}}}{r_+^{\frac{d-2}{2q-1} -
 1}}z^{\frac{d-2}{2q-1} - 1}.
\end{eqnarray}
On the boundary ($z=0$), Eq.~(\ref{phiz}) implies that
$\phi(z=0)=\mu$. According to \cite{PMb}, we have
\begin{eqnarray}\label{mu}
\mu = \frac{\rho^{\frac{1}{2q-1}}}{r_+^{\frac{d-2}{2q-1}-1}}.
\end{eqnarray}
Combining Eqs.~(\ref{T}), (\ref{Tc}), (\ref{phiz}) and (\ref{mu}),
we can find
\begin{eqnarray}\label{phizero}
\phi(z=0)=A
\frac{T_{c}^{\frac{d-2}{2q-1}}}{T^{\frac{d-2}{2q-1}-1}}.
\end{eqnarray}
We note that Eq.~(\ref{phisolsmallpsi}) is
valid for all $z$ including $z=0$. Therefore by equating
Eqs.~(\ref{phizero}) and (\ref{phisolsmallpsi}) at the boundary ($z=0$)
to find
\begin{eqnarray}\label{eqtc}
A \frac{T_{c}^{\frac{d-2}{2q-1}}}{T^{\frac{d-2}{2q-1}-1}}=
AT_{c}-(A T_{c})^{3-2q}\left[\frac
{r_{+}^{2q-2\lambda_i-2}<\mathcal{O}_i>^2}{(-1)^{3q}2^qb
q(2q-1)} \right]\chi (0).
\end{eqnarray}
First of all, we note that the
solutions of Eq.~(\ref{chi}) for $\chi(z)$
depend on the parameters $d$, $q$ and $\lambda_i$ but are
independent of $r_{+}$ , $T_{c}$ and $T$. Eq.~(\ref{chi}) may not
have an analytical solution for all values of parameters $q$, $d$
and $\lambda_{i}$ but we only need $\chi|_{z\rightarrow0}$ which
can be found numerically using the boundary conditions
$\chi(1)=\chi'(1)=0$ \cite{SD}. Thus, $\chi(z=0)$ is just a constant
which does not include $r_{+}$, $T_{c}$ or $T$. Solving Eq.~(\ref{eqtc})
for $<\mathcal{O}_i>$, after using the fact that
$\chi(0)$ is a constant, one finds
\begin{eqnarray} \label{Atc}
AT_c\left[\left(\frac{T_c}{T}\right)^{\frac{d-2}{2q-1}-1}-1\right]=s
T_c^{3-2q}T^{2q-2\lambda_i-2}<\mathcal{O}_i>^2,
\end{eqnarray}
where
\begin{eqnarray}
s=\frac{A^{3-2q} \chi(0)}{(-1)^{3q-1}2^qb
q(2q-1)}\left(\frac{4\pi}{d-1}\right)^{2q-2\lambda-2},
\end{eqnarray}
is a constant independent of $r_+$ and $T_c$. Eq.~(\ref{Atc}) can
be rearranged as
\begin{eqnarray}\label{ex}
A\left[\left(\frac{T_c}{T}\right)^{\frac{d-2}{2q-1}-1}-1\right]=s\left(\frac{T_c}{T}\right)
^{2\lambda_i-2q-2}\frac{<\mathcal{O}_i>^2}{T_c^{2\lambda_i}}.
\end{eqnarray}
Finally, the explicit expression for $<\mathcal{O}_i>$ is obtained
as
\begin{eqnarray}\label{OP}
<\mathcal{O}_i>=\tilde{A}T_c^{\lambda_i}\left(\frac{T}{T_c}
\right)^{\lambda_i-q-1}\sqrt{\left(\frac{T_c}{T}\right)^
{\frac{d-2}{2q-1}-1}
\left[1-\left(\frac{T}{T_c}\right)^{\frac{d-2}{2q-1}-1}\right]},
\end{eqnarray}
where $\tilde{A}={A}/{s}$. Near the critical temperature $T\approx
T_c$, we can write $\frac{T}{T_c}\approx\frac{T_c}{T} \approx1$,
and then $<\mathcal{O}_i>$ reduces to
\begin{eqnarray}\label{critexp}
<\mathcal{O}_i> \approx \tilde{A} T_c^{\lambda_i}
\sqrt{1-\left(\frac{T}{T_{c}}\right)^{\frac{d-2}{2q-1}-1}}.
\end{eqnarray}
Using the fact that near the critical temperature, $t=1-T/T_c$ is
very small, we can rewrite the above as
\begin{eqnarray}\label{criticalexponent}
<\mathcal{O}_i> \approx \tilde{A} T_c^{\lambda_i}
\sqrt{1-\left[1-\left(\frac{d-2}{2q-1}-1\right)t\right]}\approx
\tilde{A} T_c^{\lambda_i}\sqrt{\left(\frac{d-2}{2q-1}-1\right)t},
\end{eqnarray}
It is worth noting that the result (\ref{criticalexponent}) holds
for both the scalar operators $<\mathcal{O}_{+}>$ and
$<\mathcal{O}_{-}>$ in the allowed ranges of the Power-Maxwell
parameter $1/2<q<(d-1)/2$ and in arbitrary dimensions. It is
obvious from (\ref{criticalexponent}) that the critical exponent
of system is equal to $\beta=1/2$ which is consistent with the
mean-field value. This implies that neither the higher curvature
Gauss-Bonnet correction term nor the Power-Maxwell nonlinear
electrodynamics affect the critical exponent of the holographic
superconductors. This is not surprising as the critical exponent
is a universal feature of such theories. However, the
non-universal proportionality constant ($\tilde{A}
T_c^{\lambda_i}\sqrt{\left(\frac{d-2}{2q-1}-1\right)}$) depends on
$d$ as well as $q$. We note that the behavior of this
proportionally constant is somewhat suspect at $q=1/2$. This is
not an allowed value, but we can get close to it arbitrarily. The more
exact formula for order parameter, Eq. (\ref{OP}), also display
some limitation for $q \rightarrow 1/2$ and/or $T \rightarrow 0$
limit.

%%%%%%%%%%%%%%%%%%%%%%%%%%%%%%%%%%%%%%%%%%%%%%%%%%%%%%%%%%%%%%%%%%%%%%
\section{Concluding remarks}
Based on the Sturm-Liouville eigenvalue problem, we analytically
investigated the holographic $s$-wave superconductors with
Power-Maxwell electrodynamics in the background of Gauss-Bonnet
AdS spacetime. Our main results are analytic expressions for the
critical temperature and order parameter (which gives the critical
exponent) associated with the superconducting phase transition. We
provided a systematic illustration of the behavior of the critical
temperature as various parameters change. We also carried out a
numerical study of the critical temperature by using the shooting
method. We confirmed that the analytical results obtained based on
the Sturm-Liouville method are in good agreement with the
numerical results. The physically relevant range of Power-Maxwell
parameter $1/2<q<{(d-1)}/{2}$ is in focus here. Most importantly,
and in contrast to various other studies, we find that the
presence of nonlinear electrodynamics can lead to the easing of the
transition thus leading to higher critical temperatures for
sublinear Power-Maxwell parameter $q$. On the other hand, the
effect of Gauss-Bonnet coefficient $\alpha$ always lowers the
critical temperature. Also, transition is helped in higher
dimensions. We find that the critical exponent associated with the order
parameter is not effected by the main parameter under consideration here,
i.e, $q$, $\alpha$ and $d$.

Finally, we would like to mention that in this paper, we only
considered the probe limit, where the scalar and gauge fields do
not back react on the metric background. It would be interesting
if one could consider the Gauss-Bonnet holographic
superconductor from Power-Maxwell field away from the probe limit
and take the backreaction into account. One may also consider the case in which the
Gauss-Bonnet coupling $\alpha$ is negative \cite{GB neg}. These
issues are currently under investigation and the results will be
reported subsequently.
%%%%%%%%%%%%%%%%%%%%%%%%%%%%%%%%%%%%%%%%%%%%%%%%%%%%%%%%%%%%%%%%%%%%%%%
\acknowledgments{We thank Shiraz University Research Council. The
work of A.S has been supported financially by Research Institute
for Astronomy and Astrophysics of Maragha (RIAAM), Iran.}
%%%%%%%%%%%%%%%%%%%%%%%%%%%%%%%%%%%%%%%%%%%%%%%%%%%%%%%%%%%%%%%%%%%%%%%

\end{document}